\documentclass[%
 reprint,
 amsmath,amssymb,
 aps,
]{revtex4-2}

\usepackage{graphicx}
\usepackage{dcolumn}
\usepackage{bm}

\begin{document}

\preprint{APS/123-QED}

\title{Deflection of
barium monofluoride 
molecules using 
the bichromatic force: 
A 
density-matrix
simulation}
\author{A. Marsman}
\author{M. Horbatsch}
\author{E. A. Hessels}
 \email{hessels@yorku.ca}

\affiliation{%
 Department of Physics and Astronomy,
 York University
}%

\collaboration{EDM$^3$
Collaboration}

\date{\today}

\begin{abstract}
A full 
density-matrix
simulation is performed
for 
optical deflection of
a 
barium monofluoride
($^{138}$Ba$^{19}$F)
beam
using the 
bichromatic force,
which employs pairs of 
counter-propagating
laser beams 
that are offset in frequency.
We show that the force is 
sufficient to separate 
BaF
molecules 
from the other  
products generated in a
helium-buffer-gas-cooled
ablation source.
For our simulations,
the 
density-matrix 
and force equations
are numerically integrated
during the entire time that 
the molecules pass through 
a laser beam to ensure 
that effects of the evolution 
of the Doppler shift
and 
of the optical intensity 
and phase at the 
position of the molecule
are properly included.
The results of this work are compared to 
those of a deflection scheme 
(Phys. Rev. A 107, 032811 (2023))
which uses 
$\pi$ pulses 
to drive 
frequency-resolved
transitions.
This work is part of an effort
by the 
EDM$^3$
collaboration to 
measure the 
electric dipole moment 
of the electron
using 
BaF 
molecules
embedded in a 
cryogenic argon solid.
Separation of 
BaF
molecules 
will aid in producing 
a sufficiently pure solid.
\end{abstract}

\maketitle


\section{\label{sec:intro}Introduction}

The 
bichromatic
force uses 
two 
pairs of 
counter-propagating
laser beams that are
offset from each other in 
frequency.
The frequency offset 
leads to beat notes
in both directions,
and,
at the location of an 
atom or molecule,
this beating leads to 
pulses 
(offset in time) 
arriving
from each direction.
The effect of these 
pulses is to 
cause a promotion  
to an excited state,
followed by stimulated
emission back down 
to the ground state.
Bichromatic 
forces were
first demonstrated 
by Grimm, 
et al.
\cite{grimmObservation1990}.
As one quantum 
($\hbar \vec{k}$)
of momentum is imparted to 
the atom or molecule for 
each excitation and 
each stimulated emission, 
strong forces are possible.
For standard 
spontaneous-emission
optical forces
\cite{ashkin1970acceleration,schieder1972atomic},
the maximum force possible is
$\frac{1}{2}\hbar k/\tau$,
with the factor of 
one half
coming from equalized populations
between the ground and excited states
and 
$\tau$ 
(the
spontaneous-decay 
lifetime)
being the average 
wait time between 
laser excitations.
The 
bichromatic force 
allows for two quanta
of momentum being imparted
and allows for 
much shorter
wait times.

Bichromatic 
forces have
been used to manipulate
atomic beams of
Na
\cite{grimmObservation1990,ovchinnikov1993rectified},
Rb
\cite{gupta1993bichromatic,williamsMeasurement1999,WilliamsBichromatic2000},
Cs 
\cite{sodingShort1997},
and
of
metastable
He 
\cite{cashen2001strong,PartlowBichromatic2004,ChiedaBichromatic2012,corder2015laser}
and
Ar
\cite{feng2017bichromatic}.
More recently,
the use of 
bichromatic forces
on molecules has 
become a topic of interest
because of its
distinct advantage of imparting 
many quanta of momentum
per spontaneous
decay event.
For molecules, 
where population can be lost
due to branching ratios 
to other vibrational 
(or electronic)
states, 
optical forces with 
fewer spontaneous decays
require 
fewer 
repump lasers.
Schemes for manipulating 
molecular beams
with 
the bichromatic force
have been 
proposed
for 
CaF
\cite{chieda2011prospects,aldridge2016simulations},
MgF
\cite{yang2016bichromatic,dai2015efficient},
BaH
\cite{wenz2020large},
YbF
\cite{yin2018optically},
and
$^{137}$Ba$^{19}$F
\cite{kogel2021laser}.
Thus far, 
deflections have only 
been demonstrated using 
CaF
\cite{galica2018deflection}
and
SrOH
\cite{kozyryev2018coherent}.




In this work, 
we calculate 
the deflection of 
a 
BaF
beam 
due
to the 
bichromatic force.
The BaF molecule 
is being used by the 
EDM$^3$
collaboration to 
make an 
ultraprecise 
measurement
\cite{vutha2018orientation}
of the electric
dipole moment of the electron.
For the measurement,
BaF 
molecules 
produced by 
a 
buffer-gas-cooled
laser-ablation 
source
are embedded
in
solid 
argon. 
It is necessary to 
separate 
the
BaF 
molecules 
from the
other ablation products 
via a deflection
in order to produce
an uncontaminated solid.
For this purpose,
a deflection of 
approximately
3~m/s
(approximately 
1000
quanta of photon momentum)
is required. 
This deflection is much 
larger 
than
demonstrated 
\cite{kozyryev2018coherent,galica2018deflection}
deflections of 
other molecules 
using the
bichromatic 
force.

In 
Sect.~\ref{sec:densityMatrix}
we describe the 
density-matrix
equations used to determine 
our deflections.
Section~\ref{sec:results}
gives the results of our
calculations
and 
Sect.~\ref{sec:compare}
compares these results with 
the results of 
Ref.~\cite{marsman2023large},
which evaluates deflections
caused by
time-resolved 
$\pi$~pulses
driving
frequency-resolved
laser transitions.



\begin{figure}
\centering
\includegraphics[width=\linewidth]{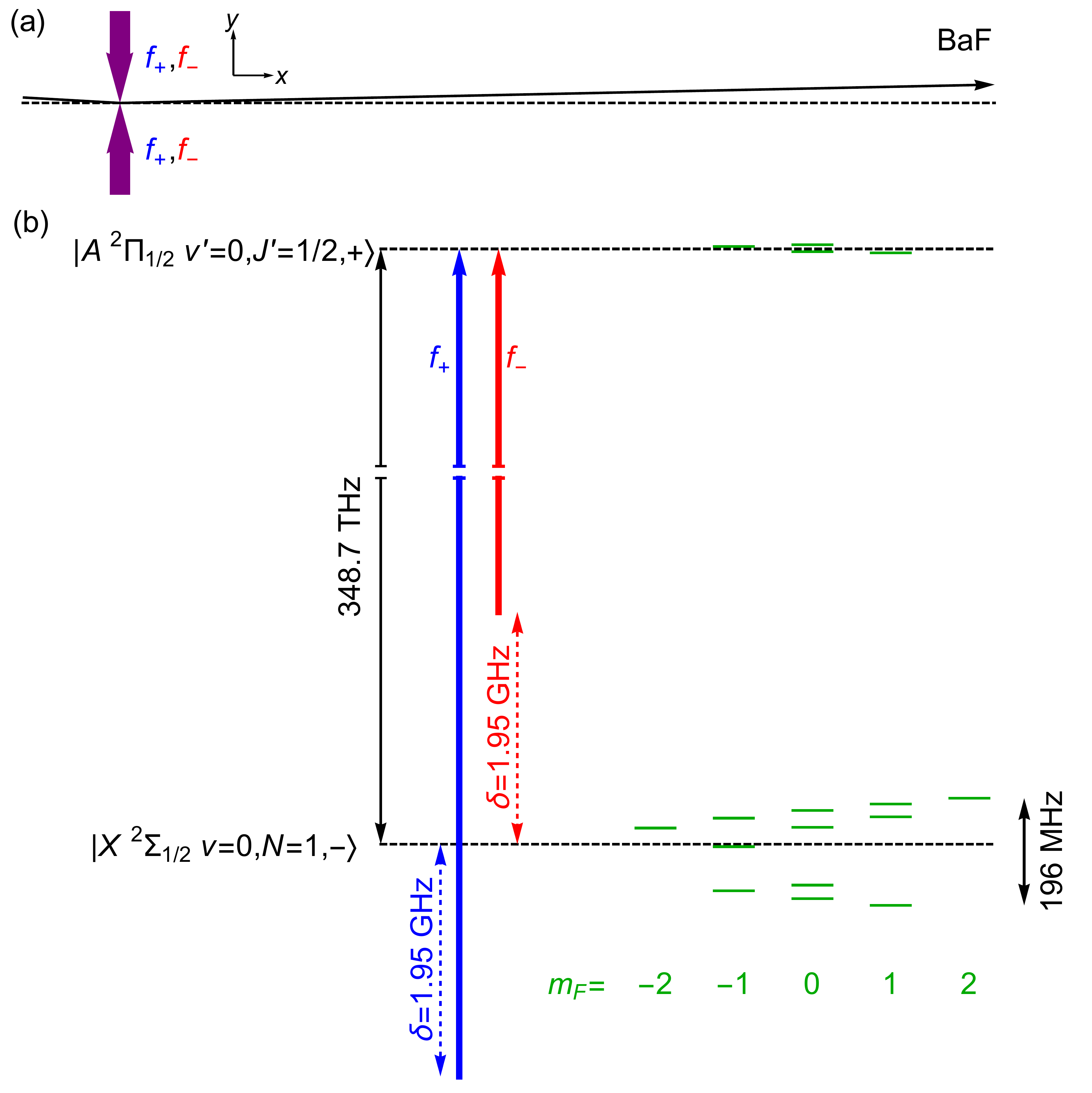}
\caption{
\label{fig:energyLevels} 
(color online)
The 
bichromatic force 
deflects
a
BaF
molecule 
using  
two pairs of 
counter-propagating 
laser beams 
(as shown in (a))
that are 
linearly polarized in the
$\hat{z}$
direction
and are
offset from each other
in frequency by 
$\pm\delta$
(as shown in (b)). 
The structure of the 
$X\,^2\Sigma_{1/2}\ v$$=$$0,\ N$$=$$1$
negative parity state
and
the
$A\,^2\Pi_{1/2}\ v'$$=$$0, j'$$=$$1/2$
positive parity state,
shown at the right on an expanded scale in 
(b),
is much smaller than 
$\delta
=
700/(2\pi\tau)
=
1950$~MHz.
A 
20-gauss 
magnetic field
(oriented $60^{\circ}$
from  
$\hat{z}$) 
lifts the degeneracy of the 
$m_F$ levels
and ensures that 
molecules do not remain
in a dark state.
}
\end{figure}

\section{Density-matrix calculation \label{sec:densityMatrix}}

In this section we present a 
density-matrix
calculation of  
BaF
molecules
being deflected 
(see 
Fig.~\ref{fig:energyLevels}(a))
by 
the
bichromatic force.
Since a
BaF 
molecule in the 
$A\,^2\Pi_{1/2}(v$$=$$0)$
state has
a branching ratio
\cite{hao2019high}
of
96.4\%
for decay to the 
$X\,^2\Sigma_{1/2}(v=0)$
state,
the 
sixteen states
of 
Fig.~\ref{fig:energyLevels}(b)
form an almost closed system,
and these sixteen states
are the only ones included in our 
density-matrix 
calculations.
Because the ground state has 
more substates than the excited
state,
spontaneous decay will
result in 
the population quickly 
landing in 
a dark state
(a linear combination 
of the
ground
states that 
is insensitive to
laser excitation).
To avoid this, 
the dark states 
are 
destabilized
\cite{berkeland2002destabilization}
by
a magnetic field 
\cite{shuman2010laser}
which causes 
the molecules 
to 
Larmor 
precess
into 
bright states.
The magnetic field used 
is
20~gauss 
in the 
$
\hat{y}\sin(60^{\circ})
+
\hat{z}\cos(60^{\circ})
$
direction of
Fig.~\ref{fig:energyLevels}(a),
but our results are found to 
be only mildly sensitive to
the exact direction and magnitude
of this field.

The 
time evolution of
the 
density matrix 
elements is determined 
by numerically
integrating their
differential equations
during the time period 
when the molecule
passes through the 
laser beams.
This time evolution is 
calculated in the 
zero-magnetic-field basis:

\begin{eqnarray}
\frac{d\rho_{g g'}}{dt}&=&
i \omega_{g'\!g} \rho_{g g'}
+
i
\sum_{g''}
(
\chi_{g'' g'} \rho_{g g''}
-
\chi_{g g''} \rho_{g'' g'}
)
\nonumber
\\
&&+
\frac{i}{2}
\sum_e 
(\Omega_{e g'} \rho_{g e}
-\Omega_{g e}  \rho_{e g'}) 
+\sum_{e,e'} \gamma_{e g e'\!g'} \rho_{e e'};
\nonumber
\end{eqnarray}
\begin{eqnarray}
\frac{d\rho_{e e'}}{dt}&=&
i \omega_{e'\! e} \rho_{e e'}
+
i
\sum_{e''}
(
\chi_{e'' e'} \rho_{e e''}
-
\chi_{e e''} \rho_{e'' e'}
)
\nonumber
\\
&&
+\frac{i}{2} \sum_{g} 
(\Omega_{g e'}\rho_{e g}
-\Omega_{e g} \rho_{g e'})
\nonumber
\\
&&-\frac{1}{2} \sum_{g,e''} 
(\gamma_{e g e''\! g} \rho_{e''\! e'}
+\gamma_{e''\! g e'\! g} \rho_{e e''});
\nonumber
\end{eqnarray}
\begin{eqnarray}
\frac{d\rho_{g e}}{dt}&=&
i(\omega_{e g}-\omega_0) \rho_{g e}
+\frac{i}{2}\sum_{g'} \Omega_{g'\!e} \rho_{g g'}
\nonumber
\\
&&+i\sum_{e'} \chi_{e' e} \rho_{g e'}
  -i\sum_{g'} \chi_{g g'} \rho_{g' e}
\nonumber
\\
&&-\frac{i}{2}\sum_{e'} \Omega_{g e' } \rho_{e'\!e}
-\frac{1}{2}\sum_{g',e'} \gamma_{e'\!g'\!eg'} \rho_{g e'},
\label{eq:denMatrix}
\end{eqnarray}
where the indices 
$g$
and 
$e$
represent the 
12
ground 
and 
4
excited states,
respectively.
Here,
$\hbar \omega_{ab}=E_a-E_b$
is the 
zero-field
energy 
difference between 
states
$|a\rangle$
and 
$|b\rangle$,
with
$\hbar \chi_{a b}=-\langle a|\vec{\mu}\cdot\vec{B}|b\rangle$
being the elements of
the 
Zeeman Hamiltonian.
The energies and 
Zeeman elements
are calculated
using the methods 
detailed in the 
appendix of 
Ref.~\cite{kaebert2021characterizing}
and are listed in 
Tables~\ref{tab:AppendixEnergies}
and
\ref{tab:AppendixZeeman}
in the appendix of this work.
The difference
between the 
average energy of 
the 
$|e\rangle$
states
and the 
average of the
$|g\rangle$
states 
is denoted by 
$\hbar \omega_{0}$.
Eq.~(\ref{eq:denMatrix})
uses the
rotating-wave 
approximation 
to avoid the 
fast oscillations 
at the optical 
frequency 
$\omega_0$.

We use the complete
formulation for spontaneous
decay
\cite{cardimona1983spontaneous,marsman2012shifts}
which includes 
quantum-mechanical 
interference 
from the decay process
using  
\begin{equation}
\gamma_{e g e'\!g'}
=
\frac{\omega^3}{3 \pi \epsilon_0 \hbar c^3}
\vec{d}_{ge} \cdot \vec{d}_{e'\!g'}.
\end{equation}
The diagonal 
elements 
$\gamma_{e g e g}$
are equal to the 
branching ratio 
times 
$1/\tau$.
The 
off-diagonal 
elements properly account for 
quantum-mechanical 
interference
and, 
to our knowledge,
this is the first time they 
have been included in 
calculations of the 
bichromatic
force.
For the present case, 
these terms lead to only
small corrections.

The dipole matrix elements
can be deduced from 
the measured lifetime of the 
$A\,^2\Pi_{1/2}$
state  
($\tau=$
57.1~ns
\cite{aggarwal2019lifetime}),
along with the 
branching ratio 
\cite{hao2019high} 
between individual 
hyperfine 
states.
These latter ratios are 
calculated using the methods
described in 
Ref.~\cite{kaebert2021characterizing}.
The values of the
electric dipole matrix elements
are listed 
in 
Table~\ref{tab:AppendixMatrixEls} in 
the 
appendix.

The 
Rabi
frequencies
in 
Eq.~(\ref{eq:denMatrix})
are given by
\begin{eqnarray}
\label{eq:fRabi}
\Omega_{eg}(t)
\!&=&\!
\frac{\vec{d}_{eg} \cdot \hat{z} E_{0}(x)}{\hbar}
\!\!\!\!\sum_{s,\sigma=\pm 1}
\!\!\!\!e^{i (s\, k\, y+2\pi\,\sigma\,\delta\, t - s\, \sigma\, 
\frac
{\pi}
{8}
+\phi)}
\Bigg\rvert_{
\substack{
(x,y,z)\\
=\vec{r}_m(t)}
},
\nonumber
\\
&& 
\end{eqnarray}
where the 
sum is 
over the four laser beams
(with frequencies offset by 
$\pm \delta$
and 
with 
$\vec{k}$
in the 
$\pm \hat{y}$
directions).
For the present work,
$\delta$
is chosen to be 
$700/(2\pi\tau)$,
or
1950~MHz.
Note our choice to 
express 
$\delta$
as a frequency, 
rather than an
angular frequency.
The field amplitude
$E_{0}(x)$
includes
the spatial profile
of the laser beam, 
which is taken to be 
a 
top-hat 
shape 
(as, 
e.g.,
in 
Ref.~\cite{ma2011improvement})
approximated by a 
super-gaussian
function
of the form
\begin{equation}
\label{eq:profile}
E_{\rm max} e^{-(x^2/w^2)^5},
\end{equation}
where the value of  
$w$
is chosen to give an
intensity profile with 
a full width at 
half-maximum
of
0.32~mm
and 
$E_{\rm max}$
is set to the 
value that gives the best 
approximation to  
$\pi$ 
pulses
for a multilevel system
\cite{aldridge2016simulations}:
$E_{\rm max}
=
2 \pi \delta \sqrt{6 \pi h \tau /(\lambda^3 \epsilon_0 )}=$
1380~V/cm 
(corresponding to 
a laser power of 
2~W for each of the 
four beams).

Both
the 
$x$
dependence 
in the profile of
Eq.~(\ref{eq:profile})
and the 
$y$ found 
in the complex
exponential of 
Eq.~(\ref{eq:fRabi})
are evaluated at the 
molecular position
$\vec{r}_m(t)$.
The 
trajectory 
of the molecule
is obtained from 
its original position,
its initial velocity
(assumed to be 
150~m/s
--
a typical speed
for a
BaF 
molecule
in a
4-kelvin
helium-buffer-gas-cooled
beam
\cite{truppe2018buffer}),
and the
force
obtained
\cite{cook1979atomic}
from
Ehrenfest's
theorem:

\begin{equation}
\vec{F}(t)=-  \hbar \sum_{e,g} {\rm Re}
[
\rho_{e g}(t) \nabla \Omega_{g e}(\vec{r},t)
]
\Bigg\rvert_{\vec{r}=\vec{r}_m(t)}.    
\end{equation}

As discussed above, 
the 
density-matrix
treatment is not complete
in that the sixteen
levels do not quite form
a closed system.
The remaining branching ratio
(3.6\%)
causes population to be lost
out of the cycling transition
and
is primarily due to 
radiative 
decay down to the 
$X\,^2\Sigma_{1/2}(v=1)$
state
(3.5\%)
\cite{hao2019high},
with much smaller 
contributions for
decays to 
the 
$X\,^2\Sigma_{1/2}(v$$>$$1)$
states 
and the 
$A'\,^2\Delta_{3/2}$
state.
Spontaneous decay to 
the latter state
causes the molecule to 
change its parity upon
its second decay down 
to the 
$X\,^2\Sigma_{1/2}$
state.
The net effect is that 
every spontaneous emission
leads to 
3.6\% of the 
remaining population 
going into a dark state
that will no longer 
experience the 
bichromatic
force.
Fortunately, 
as we will see,
the 
bichromatic 
force 
can be applied with 
very little spontaneous
emission occurring, 
and therefore
this loss remains relatively small
and no 
repumping 
into the cycling states
is required.
We treat these dark states
by reducing the population 
of the sixteen states
in accordance with the 
build up of population into 
the dark states. 
Our density matrix calculation
continues with only the 
remaining population present,
and therefore our results
apply only to the molecules
that are not lost to the dark 
state.



The 
density matrix elements,
along with the components 
of 
$\vec{r}_m(t)$
and 
$\dot{\vec{r}}_m(t)$
are numerically integrated 
over the 
$\sim$2.7~$\mu$s
that it takes for the 
molecules to pass 
through the 
laser profile.
These 
integrations 
are
computationally intensive
given the 
3.9-GHz
frequency difference between 
the two laser beams, 
which leads to fast
oscillations in the
density-matrix
elements,
and these 
oscillations must be 
calculated accurately.

\section{Results
\label{sec:results}
}

\begin{figure}
\centering
\includegraphics[width=3.5in]{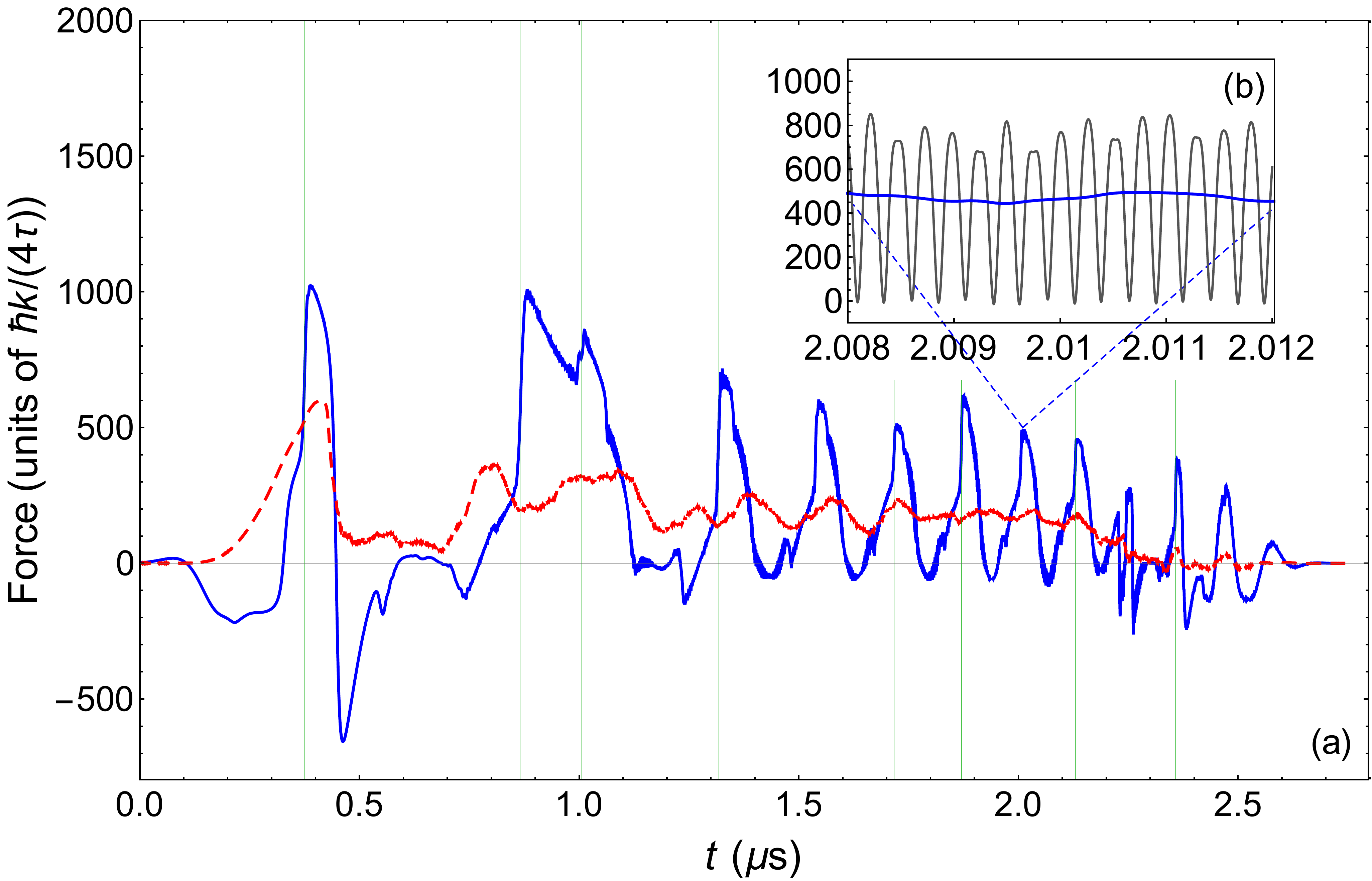}
\caption{
\label{fig:force} 
(color online)
The bichromatic
force on the 
BaF
molecule 
(with initial velocity of 
$v_y^{\rm init}=-1$~m/s)
as it passes through 
the bichromatic laser beams.
The force oscillates
at the beat frequency
(of 
$2\times1950$~MHz),
as shown in the inset (b).
The main panel (a)
shows the force averaged 
over these fast oscillations.
The blue solid curve is the force 
for a molecule initially on axis,
and the peaks versus time 
occur each time the molecule moves
by one half of a wavelength 
($\lambda/2$),
where the fields of the two beams interfere
constructively
(the times at which the molecule 
is at these positions are indicated 
by green vertical lines).
The red dashed curve is 
the force averaged over 
initial position 
($0\le y^{\rm init} < \lambda/2$).
A large average force 
of over 
$200\hbar k/(4\tau)$
is calculated.
} 
\end{figure} 

We start by considering
a molecule that is initially at 
$y^{\rm init}=0$ 
(on axis, 
see 
Fig.~\ref{fig:energyLevels})
with a small downward 
velocity of
$v_y^{\rm init}=-1$~m/s.
The force on 
the 
BaF
molecule
as it traverses the 
bichromatic laser beams 
is determined  
from our numerical 
integration 
and
is shown 
in 
Fig.~\ref{fig:force}.
The inset in the figure
shows that the force 
oscillates at the beat 
frequency
($2\times1950$~MHz),
and the forces shown in 
Fig.~\ref{fig:force}(a)
are the averages over these
fast oscillations.
The blue solid curve 
shows the force versus
time for initial conditions 
$v_y^{\rm init}=-1$,
and
$y^{\rm init}=0$.
The force has a peak value
each time the molecule is 
displaced by a half wavelength,
showing that there is a 
strong dependence on the relative
phases of the upward and downward
directed beams.
The force is large when
the fields are interfering  
constructively,
and small when they
are interfering 
destructively.
Because of this dependence, 
the force also depends strongly 
on 
$y^{\rm init}$
as this value is varied over 
half of the laser wavelength, 
$\lambda$.
Since our molecular beam has a much
larger extent
than 
$\lambda$
along 
$y$,
we average the force 
over this initial position
(averaging over
$y^{\rm init}$
that span 
$\lambda/2$),
as shown by the red dashed
curve in 
Fig.~\ref{fig:force}.

Note that the 
averaged calculated force is 
more than
200
times larger than the 
spontaneous-emission 
optical force of 
$\frac{1}{4}\hbar k/\tau$,
where here the usual
factor of one half for a 
two-level
system is replaced
by one quarter since now 
only one quarter of our sixteen
states are excited states.
This force can be compared to 
a simplified model for 
the  
bichromatic
force
in a 
two-level 
system,
in which  
the laser fields
act as a series of 
$\pi$
pulses,
with 
each period
$T=1/(2\delta)$
of the 
beat-frequency having 
one pulse from 
the 
upward-directed
laser 
followed 
a time 
$T/4$
later
by a pulse from the 
downward-directed
laser.
Ideally, 
the first pulse causes
a laser excitation and the 
second a stimulated emission
back to the ground state,
leading to a force of 
$2\hbar k/T$.
However, 
a spontaneous emission 
(if it occurs in the 
time interval between the first and 
second 
pulse) 
can reverse this 
sequence, 
leading to a force
of 
$-2\hbar k/T$.
Fortunately, 
another spontaneous emission
(if it occurs in the 
$3T/4$
interval after the second
pulse) 
again reverses the force, 
and since this time interval is three
times larger, 
it is more likely to occur.
As a result,
this simple model predicts 
a force of 
$\frac{3}{4}2\hbar k/T$
$-$
$\frac{1}{4}2\hbar k/T$
$=$
$\hbar k/T$.
This estimate is approximately 
a factor of four larger than 
our calculated average force
of
$200\frac{1}{4} \hbar k/\tau = 0.23 \hbar k/T$.
Part of this factor of four
results from the reduced effectiveness
of the bichromatic force for  
a multilevel system such as the one shown
in 
Fig.~\ref{fig:energyLevels}(b).

The first reason for reduced 
effectiveness is that
the 
number of 
substates
of the 
ground state is larger than 
that of the excited state
(the excited states
represent 
1/4
rather
than 
1/2
of the states),
which reduces the 
fraction of molecules
that can participate
at any time.
Second,
as discussed above, 
dark states would make the 
molecules insensitive to the 
lasers, reducing the 
force to zero. 
The
application of a 
magnetic field
avoids the zeroing of the 
force, 
but a reduction is still 
seen.
Finally, 
because each
$g$
state 
has a different dipole matrix
element
connecting it to the excited
states 
(and, 
to a smaller extent,
because each has a different resonant
frequency) 
it is not possible to perfectly meet the 
$\pi$-pulse 
condition for all of the 
ground 
substates.

These
factors,
along with effects that are 
already present in the 
two-level system
(such as the interference
between the fields from 
the two directions),
account for 
the factor of 
four reduction 
in 
force 
compared to 
the simple 
model.
This factor is
similar to 
the reduction calculated in other works
\cite{aldridge2016simulations,
kozyryev2018coherent}.

\begin{figure}
\centering
\includegraphics[width=3.5in]{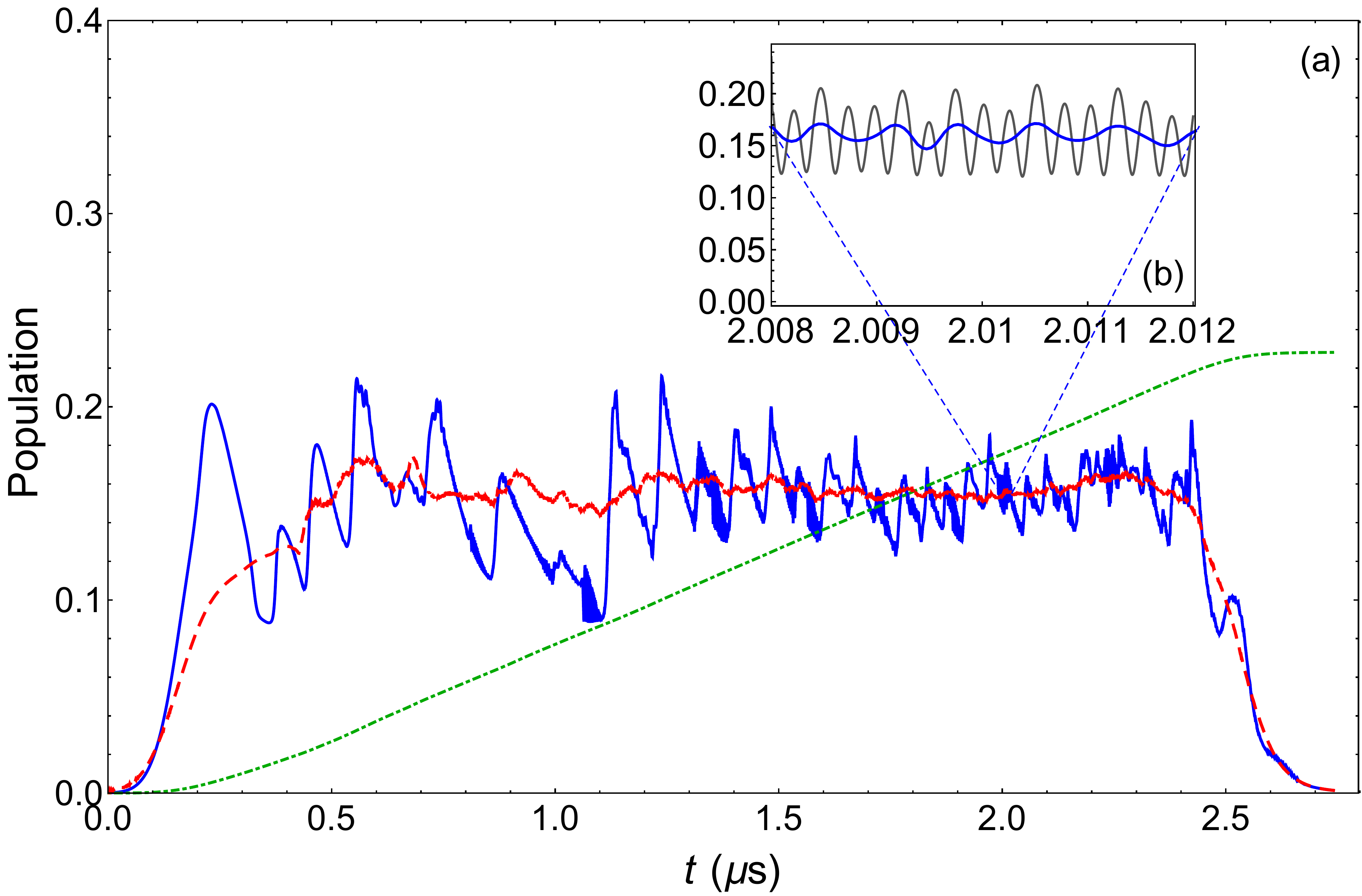}
\caption{
\label{fig:populations} 
(color online)
The 
total populations of the 
excited and dark states
for 
$v_y^{\rm init}=-1$~m/s.
Approximately 
15\%
of the molecules are
in the excited state
(blue) 
during the time that they pass 
through the laser beam.
The inset 
(b)
shows that 
this
population oscillates 
at the 
3.9-GHz
beat frequency,
as the population is 
excited up followed
by
stimulated emission 
back to the ground state.
The main plot 
(a) 
gives a 
time-averaged
view
that averages over
these oscillations. 
The blue solid line
is for 
$y^{\rm init}=0$
and the red dashed curve
is averaged over 
$y^{\rm init}$.
The population of the 
dark state 
(green dash-dot)
steadily 
increases to approximately
23\% 
as the molecules pass through 
the laser, 
allowing the remaining
77\% 
of the molecules
to experience the full force
shown in 
Fig.~\ref{fig:force}.
} 
\end{figure} 

Figure~\ref{fig:populations}
shows the time progression of 
the total 
excited-state
and 
dark-state
populations
for  
$v_y^{\rm init}=-1$~m/s.
Note that the 
population in the dark state
builds up to approximately
23\% 
by the time the molecules
have passed through 
the beam, 
allowing the majority of the 
molecules to experience
the full deflection force
of
Fig.~\ref{fig:force}.

The population in the 
excited state is
approximately 
15\%
on average 
during the time
the molecules spend 
in the laser beams.
This
15\% 
population
implies that there are only 
approximately six
spontaneous emission events
per molecule
during the 
time
it takes to pass through the 
0.32-mm 
laser beams,
whereas the molecule
has received approximately
1600 
quanta 
($1600\, \hbar k$)
of momentum kicks 
from the lasers.

The inset in 
Fig.~\ref{fig:populations}
shows that the 
excited-state 
population
oscillates at the 
3.9-GHz
beat frequency,
as the population is excited
by one laser followed by 
stimulated emission from the 
oppositely directed laser beam.
The figure shows the 
excited-state population for 
$y^{\rm init}=0$
(blue),
as well as the 
result averaged over 
$y^{\rm init}$
(red dashed).

\begin{figure}
\centering
\includegraphics[width=3.1in]{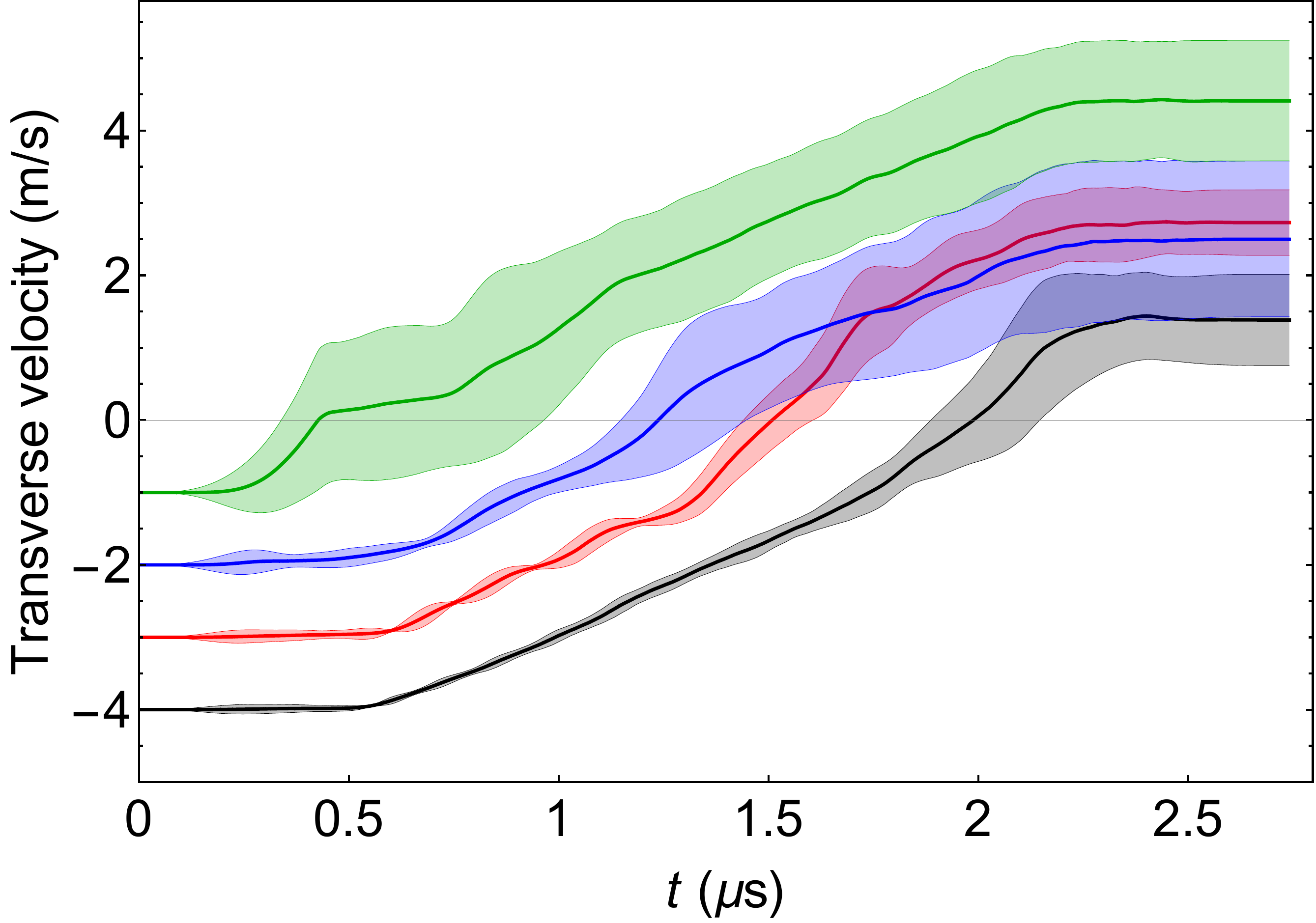}
\caption{
\label{fig:velVsTime} 
(color online)
The transverse velocity
($v_y$)
versus time 
(while passing 
through the 
bichromatic
laser beams)
for 
BaF
molecules starting with
$v_x=150$~m/s and
$v_y^{\rm init}=$
$-4$,
$-3$,
$-2$,
and
$-1$~m/s.
The molecules are 
deflected upward by 
approximately 
5~m/s.
} 
\end{figure}

Results for different 
$v_y^{\rm init}$
are illustrated 
in 
Fig.~\ref{fig:velVsTime},
where the transverse 
velocity
($v_y$)
is plotted 
versus time for 
$v_y^{\rm init}=$
$-1$,
$-2$,
$-3$,
and
$-4$~m/s.
These curves depend 
on 
$y^{\rm init}$,
and the figure shows
the averages
over 
$y^{\rm init}$,
as well as the 
one-standard-deviation 
range for variations
from this average value.
In each case, 
a deflection of approximately
5~m/s
changes the transverse 
velocity into the positive
direction.

\begin{figure}
\centering
\includegraphics[width=3.5in]{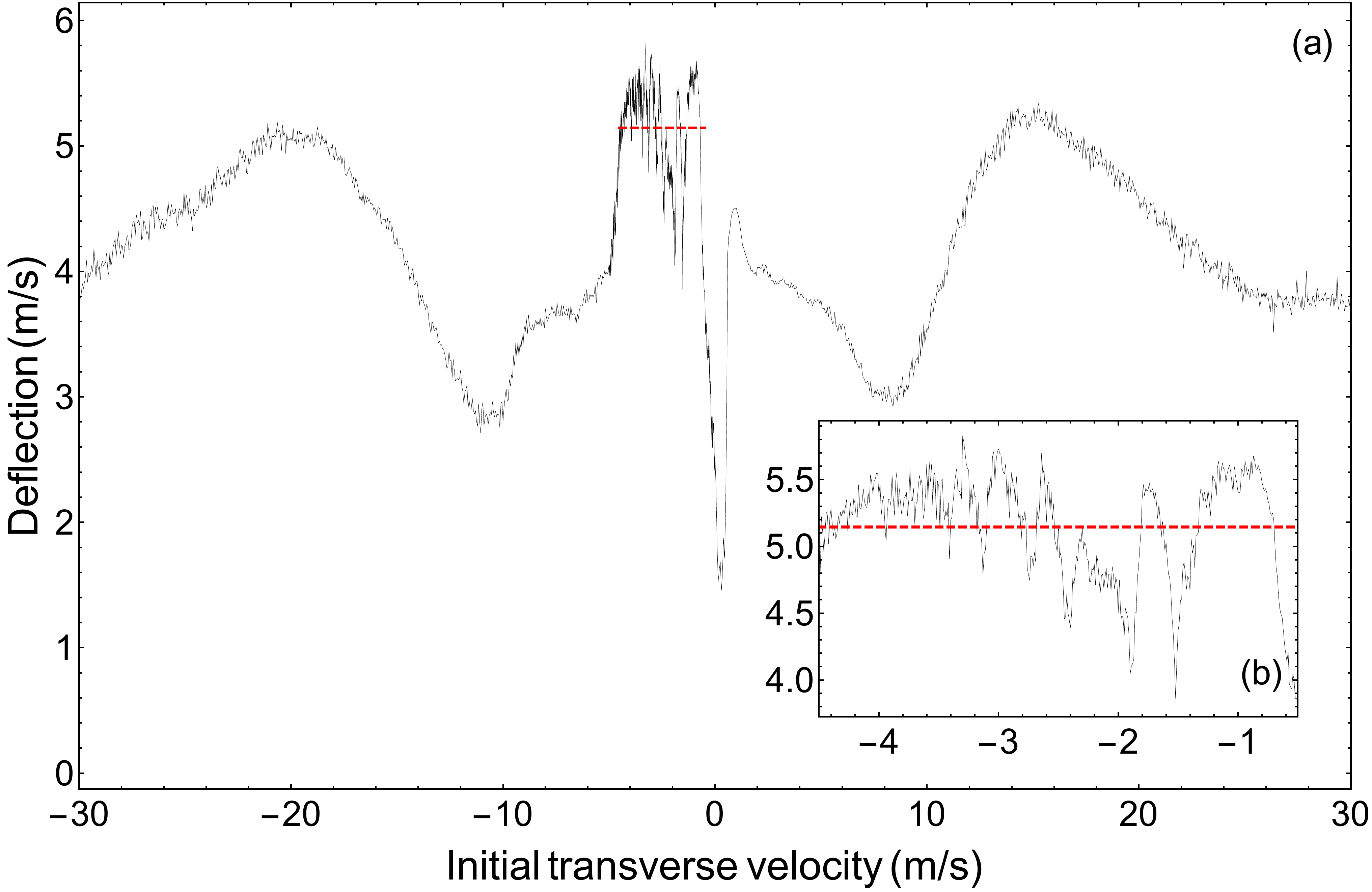}
\caption{
\label{fig:deflProfile} 
(color online)
The deflection of a
BaF
beam 
(with a longitudinal speed $v_x=150$~m/s)
after passing  
through the 
bichromatic
laser beams
as a function of its
initial transverse velocity
($v_y$).
Note that large deflections
occur over a wide
range of initial 
$v_y$.
For the present 
work, 
we focus on 
initial transverse
velocities in the 
range of 
$-4.5$
to
-0.5~m/s
(inset),
where
the average 
deflection 
is 
5.1~m/s
(red dashed line).
} 
\end{figure}

For the present work, 
the deflection is only 
needed for the small range 
of 
$v_y^{\rm init}$
shown in 
Fig.~\ref{fig:velVsTime},
but 
Fig.~\ref{fig:deflProfile}
shows that large 
deflections persist for 
a wide range of initial
transverse velocities.
Again an average over 
a half wavelength for
$y^{\rm init}$
is used for this figure.
As expected,
the large value of 
$\delta$
leads to a large force
over a wide range of transverse
velocities.

\section{comparison to 
deflection scheme using 
time-resolved 
$\pi$ pulses
\label{sec:compare}
}

\begin{table*}[]
\begin{ruledtabular}
\centering
\begin{tabular}{llccc}
Quantity&
units&
bichromatic&
bichromatic&
$\pi$ pulses
\\
&
&
this work&
expanded beam&
Ref.
\cite{marsman2023large}
\\
\hline
momentum change, $\Delta p_y$, per lifetime&
$\hbar k$&
50&
27&
2
\\
period (beat period or cycle period)&
ns&
0.26&
0.45&
10
\\
momentum change, $\Delta p_y$, per period&
$\hbar k$&
0.23&
0.21&
0.35
\\
number of frequencies:
main laser&
&
2&
2&
6
\\
\ \ \ \ \ \ \ \ \ \ \ \ \ \ \ \ \ \ \ \ \ \ \ \ \ \ \ \ \ \ \ 
repump laser&
&
0&
0&
6
\\
power per beam at each frequency&
W&
2&
2&
$\sim$2
\\
laser field
$E_0$
per beam at each frequency&
V/cm&
1380&
780&
$\sim$200
\\
laser intensity FWHM: 
$z$ direction of 
Fig.~\ref{fig:energyLevels}(a)&
mm&
0.32&
1&
1
\\
\ \ \ \ \ \ \ \ \ \ \ \ \ \ \ \ \ \ \ \ \ \ \ \ \ \ \ \ \ \ \ 
$x$ direction of 
Fig.~\ref{fig:energyLevels}(a)&
mm&
0.32&
0.32&
5
\\
momentum change, $\Delta p_y$, per spontaneous decay&
$\hbar k$&
270&
140&
20
\\
applied magnetic field&
gauss&
20&
20&
1000
\\
change in $\Delta v_y$ for imperfect laser parameters:
\\
\ \ $E_0$ for each laser varied by factor specified&
&
[0.99-1.01]:10\%
&
15\%
&
[0.95-1.05]:2\%
\\
\ \ relative phases varied by angle range specified& 

&
[0-25~mrad]:10\%
&
16\%
&
[0-2$\pi$]:6\%
\\
\ \ each laser includes 0-1\% of other polarization
&
&
[0-1\%]:6\%
&
5\%
&
[0-1\%]1\%
\\
deflection, $\Delta v_y$&
m/s&
5.1&
2.7&
2.6
\end{tabular}
\end{ruledtabular}
\caption{
A comparison of 
the parameters and results of 
this work to those of 
Ref.~\cite{marsman2023large}.
}
\label{tab:ComparePrev}
\end{table*}

The results of the present 
work can be compared to 
a recent work
\cite{marsman2023large}
which calculated deflections
of
BaF
molecules
caused by 
short-duration
$\pi$~pulses
in a large magnetic field
which resolved individual
laser transitions.
Table~\ref{tab:ComparePrev}
compares the parameters and 
results of that work to the 
present work.
The 
time-resolved
$\pi$~pulses
and 
frequency-resolved 
laser transitions
in 
Ref.~\cite{marsman2023large}
had the advantages of: 
(i)
avoiding interference 
between the fields 
from the 
oppositely-directed
lasers,
(ii)
being able to 
tune the individual 
laser powers to meet
the 
$\pi$-pulse 
condition for 
each ground state,
and 
(iii)
being able to 
adjust the timing
to maximize the net 
force.
These advantages can be
seen in the third row 
of 
Table~\ref{tab:ComparePrev},
where 
Ref.~\cite{marsman2023large}
has a 
50\% larger momentum imparted
per period
(per 
10-ns
period for the pair of 
$\pi$
pulses
in 
Ref.~\cite{marsman2023large}
and 
per period of the 
beat note in this work).
In both works,
the power 
(at each frequency)
per laser beam
is approximately 
2~W.
However, 
the extent of the 
laser beam in 
the present
work 
is 
0.32~mm
in the 
$z$ direction
(see 
Fig.~\ref{fig:energyLevels}(a)), 
whereas in 
Ref.~\cite{marsman2023large}
the extent was 
1~mm,
which allowed for interaction 
with 
three times the number of 
BaF
molecules.

To make a more direct 
comparison to the previous work
\cite{marsman2023large},
we also calculated 
bichromatic 
deflections with 
a 
2-W laser beam that has a 
FWHM 
of 
1~mm
in the 
$z$
direction, 
and these results are 
shown in 
column~4
of 
Table~\ref{tab:ComparePrev}.
The results in this column
show that the 
bichromatic
force is, 
in most respects,
a better choice for 
deflection compared to 
the 
$\pi$-pulse scheme.
The main advantage comes from the 
rate at which 
pairs of laser 
pulses 
are incident on the molecules,
which is more than a factor
of 
20
faster
(see 
row~2)
for the 
bichromatic 
case.
Even with the lower
efficiency per period,
the net force
(see 
row~1) 
is still more than 
an order of magnitude
larger.
More importantly,
the smaller period
allows for more momentum
transfer per 
spontaneous-emission
event
(see
row~10).
For both columns 
3
and 
4,
spontaneous emission 
is rare enough that
less than 
25\%
of the population 
is lost to a dark
state, 
and therefore no 
repump lasers
are needed.

In principle, 
the 
$\pi$-pulse 
scheme could also 
eliminate the need for 
repump lasers 
by having a much 
faster rate of 
$\pi$
pulses,
but this would considerably
complicate the 
scheme since the shorter
pulses would require
higher laser power,
as well as a larger magnetic
field to resolve the 
individual transitions
(as discussed in 
Ref.~\cite{marsman2023large}).

One way in which the 
$\pi$-pulse
scheme is superior
is in the robustness
of the force.
In 
Ref.~\cite{marsman2023large},
the deflection is found to vary
by only 
2\%
when the electric field
of each of the laser beams
is randomly varied between
0.95 
and
1.05
of their 
nominal values.
Here, 
by comparison,
(as shown in 
row~12
of 
Table~\ref{tab:ComparePrev}),
the deflection
varies by 
16\% 
when the relative
amplitudes (electric fields)
for the four lasers are 
randomly varied in the 
more limited range 
of 
0.99
to 
1.01.
For 
Ref.~\cite{marsman2023large},
variation of the relative
phase of the laser beam
(randomly selected over 
the full range of 
0
to
2$\pi$)
led to a 
variation in the 
deflection of 
6\%.
For the bichromatic force,
the relative phases need 
to be fixed to get a net
force, 
and we find that varying
the relative phases
randomly within
the range of 
$0$
to
$0.025$
radians
leads to a variation 
in deflection of 
15\%.
Finally, 
in 
Ref.~\cite{marsman2023large}
having imperfect polarization
(i.e., an admixture of the 
opposite circular polarization)
randomly picked between 
0
and
1\%
led to 
a variation in deflection of 
1\%.
Here, 
a similar level of imperfect polarization
(with an admixture of 
polarization along the 
orthogonal
$\hat{x}$
direction)
leads to a variation in 
deflection of 
5\%.

The final row 
of
Table~\ref{tab:ComparePrev}
shows that 
the deflection from 
the 
laser beams of
Ref.~\cite{marsman2023large}
is
nearly equal to that from
the
bichromatic 
laser beams 
(column~4). 
Although the 
bichromatic deflection scheme
does require better control
of the laser parameters, 
it is still far
simpler to implement
in that it only requires 
two 
(compared to twelve)
laser frequencies
and it requires a much smaller
magnetic field.

\section{\label{sec:concl}
Conclusions}

In this work, 
we have performed a complete
density-matrix
simulation of 
deflections of 
BaF 
molecules using the 
bichromatic force.
This deflection would allow for
a beam 
of 
$^{138}$Ba$^{19}$F
to be separated from 
other laser ablation 
products coming from 
a
buffer-gas-cooled
laser-ablation
source,
as required by the 
EDM$^3$
collaboration 
for their planned measurement 
of the electric dipole moment
of the electron 
using 
BaF
molecules embedded in an 
Ar 
solid.


\section*{\label{sec:ackn}
Acknowledgements}

We acknowledge support from
the 
Gordon and Betty Moore Foundation,
the
Alfred P. Sloan Foundation,
the
John Templeton Foundation 
(through 
the 
Center for Fundamental Physics
at
Northwestern University),
the 
Natural Sciences
and Engineering Council 
of Canada, 
the 
Canada Foundation for Innovation, 
the
Ontario Research Fund
and
from
York University.
Computations for this work were 
enabled by support provided by 
the 
Digital Research Alliance of Canada,
Compute Ontario and 
SHARCNET.

\section*{Appendix}
The energies,
Zeeman structure
and 
dipole matrix elements
for our
system were calculated using the 
methods described in 
Ref.~\cite{kaebert2021characterizing}.
The 
energies 
(in zero magnetic field)
are shown in 
Table~\ref{tab:AppendixEnergies},
the dipole matrix elements
are shown in 
Table~\ref{tab:AppendixMatrixEls},
and the 
Zeeman
matrix elements 
are shown in 
Table~\ref{tab:AppendixZeeman}.

 \begin{table}[]
 \begin{ruledtabular}
 \centering
 \begin{tabular}{cccc}
 state&$F$&$m_F$&$E/h$(MHz)\\
 \hline
$g1$&$1$&$-1$&$-94.947$\\
$g2$&$1$&$0$&$-94.947$\\
$g3$&$1$&$+1$&$-94.947$\\
$g4$&$0$&$0$&$-67.134$\\
$g5$&$1$&$-1$&$22.715$\\
$g6$&$1$&$0$&$22.715$\\
$g7$&$1$&$+1$&$22.715$\\
$g8$&$2$&$-2$&$56.766$\\
$g9$&$2$&$-1$&$56.766$\\
$g10$&$2$&$0$&$56.766$\\
$g11$&$2$&$+1$&$56.766$\\
$g12$&$2$&$+2$&$56.766$\\
$e1$&$1$&$-1$&$\omega_0/(2\pi)-1.25$\\
$e2$&$1$&$0$&$\omega_0/(2\pi)-1.25$\\
$e3$&$1$&$+1$&$\omega_0/(2\pi)-1.25$\\
$e4$&$0$&$0$&$\omega_0/(2\pi)+3.75$
\end{tabular}
\end{ruledtabular}
\caption{
The energies 
of the ground states
and 
excited states
in zero magnetic field.
}
\label{tab:AppendixEnergies}
\end{table}

\begin{table}[]
\begin{ruledtabular}
\centering
\begin{tabular}{cccc}
ground&excited&polar- &$d_{eg}$\\
state &state  &ization&($ea_0$)\\ 
\hline
$g1$&$e1$&$0$&$-1.17$\\
$g1$&$e2$&$+1$&$1.17$\\
$g1$&$e4$&$+1$&$0.839$\\
$g2$&$e1$&$-1$&$1.17$\\
$g2$&$e3$&$+1$&$1.17$\\
$g2$&$e4$&$0$&$-0.839$\\
$g3$&$e2$&$-1$&$-1.17$\\
$g3$&$e3$&$0$&$-1.17$\\
$g3$&$e4$&$-1$&$0.839$\\
$g4$&$e1$&$-1$&$1.105$\\
$g4$&$e2$&$0$&$-1.105$\\
$g4$&$e3$&$+1$&$-1.105$\\
$g5$&$e1$&$0$&$-0.063$\\
$g5$&$e2$&$+1$&$0.063$\\
$g5$&$e4$&$+1$&$-1.062$\\
$g6$&$e1$&$-1$&$-0.063$\\
$g6$&$e3$&$+1$&$-0.063$\\
$g6$&$e4$&$0$&$-1.062$\\
$g7$&$e2$&$-1$&$-0.063$\\
$g7$&$e3$&$0$&$-0.063$\\
$g7$&$e4$&$-1$&$-1.062$\\
$g8$&$e1$&$+1$&$-0.957$\\
$g9$&$e1$&$0$&$0.677$\\
$g9$&$e2$&$+1$&$0.677$\\
$g10$&$e1$&$-1$&$-0.391$\\
$g10$&$e2$&$0$&$-0.781$\\
$g10$&$e3$&$+1$&$0.391$\\
$g11$&$e2$&$-1$&$0.677$\\
$g11$&$e3$&$0$&$-0.677$\\
$g12$&$e3$&$-1$&$-0.957$
\end{tabular}
\end{ruledtabular}
\caption{
The 
electric-dipole
matrix
elements between 
the states 
of 
Table~\ref{tab:AppendixEnergies}.
}
\label{tab:AppendixMatrixEls}
\end{table}

\begin{table}[]
\begin{ruledtabular}
\centering
\begin{tabular}{cccc}
state 1 &state 2  &component, $q$&$\langle 1|\mu_q|2\rangle (\mu_B)$\\ 
\hline
$g1$&$g1$&$0$&$0.498764$\\
$g1$&$g2$&$-1$&$-0.508397$\\
$g2$&$g3$&$-1$&$-0.508397$\\
$g3$&$g3$&$0$&$-0.498765$\\
$g4$&$g1$&$1$&$0.0611559$\\
$g4$&$g2$&$0$&$-0.0540664$\\
$g4$&$g3$&$-1$&$0.0611559$\\
$g5$&$g1$&$0$&$0.0861576$\\
$g5$&$g2$&$-1$&$-0.0838738$\\
$g5$&$g4$&$-1$&$1.00933$\\
$g5$&$g5$&$0$&$-0.976845$\\
$g6$&$g1$&$1$&$-0.0838738$\\
$g6$&$g3$&$-1$&$0.0838738$\\
$g6$&$g4$&$0$&$-1.00037$\\
$g6$&$g5$&$1$&$0.986477$\\
$g7$&$g2$&$1$&$0.0838738$\\
$g7$&$g3$&$0$&$-0.0861576$\\
$g7$&$g4$&$1$&$1.00933$\\
$g7$&$g6$&$1$&$0.986477$\\
$g7$&$g7$&$0$&$0.976845$\\
$g8$&$g1$&$-1$&$-1.23099$\\
$g8$&$g5$&$-1$&$-0.0572771$\\
$g8$&$g8$&$0$&$-0.97496$\\
$g9$&$g1$&$0$&$-0.864044$\\
$g9$&$g2$&$-1$&$-0.87044$\\
$g9$&$g5$&$0$&$-0.0390138$\\
$g9$&$g6$&$-1$&$0.040501$\\
$g9$&$g8$&$1$&$-0.69732$\\
$g9$&$g9$&$0$&$-0.48748$\\
$g10$&$g1$&$1$&$-0.502549$\\
$g10$&$g2$&$0$&$-0.997712$\\
$g10$&$g3$&$-1$&$-0.502549$\\
$g10$&$g5$&$1$&$-0.0233833$\\
$g10$&$g6$&$0$&$0.0450493$\\
$g10$&$g7$&$-1$&$-0.0233833$\\
$g10$&$g9$&$1$&$-0.85404$\\
$g11$&$g10$&$1$&$-0.85404$\\
$g11$&$g11$&$0$&$0.48748$\\
$g11$&$g2$&$1$&$-0.87044$\\
$g11$&$g3$&$0$&$-0.864044$\\
$g11$&$g6$&$1$&$0.040501$\\
$g11$&$g7$&$0$&$-0.0390138$\\
$g12$&$g11$&$1$&$0.69732$\\
$g12$&$g12$&$0$&$0.97496$\\
$g12$&$g3$&$1$&$1.23099$\\
$g12$&$g7$&$1$&$0.0572771$\\
$e1$&$e1$&$0$&$0.20784$\\
$e3$&$e3$&$0$&$-0.20784$\\
$e4$&$e2$&$0$&$-0.20784$
\end{tabular}
\end{ruledtabular}
\caption{
The 
nonzero
magnetic dipole 
matrix
elements
between 
the states 
of 
Table~\ref{tab:AppendixEnergies}
(in spherical tensor form).
}
\label{tab:AppendixZeeman}
\end{table}

\bibliography{bichromaticBaF}

\end{document}